\documentclass[twocolumn,aps,prd,floatfix,nofootinbib,superscriptaddress]{revtex4}
\usepackage{epsfig}
\usepackage{graphicx}
\usepackage{amssymb,amsmath}
\usepackage{pdflscape}

\begin{document}

\title{\boldmath Measurement of the neutron timelike
electric and magnetic form factors ratio at the VEPP-2000 $e^+e^-$ collider}

\newcommand{\binp}{\affiliation{Budker Institute of Nuclear Physics,
SB RAS, Novosibirsk 630090, Russia}}
\newcommand{\nsu}{\affiliation{Novosibirsk State University,
Novosibirsk 630090, Russia}}

\author{M.~N.~Achasov} \binp\nsu
\author{A.~E.~Alizzi} \binp\nsu
\author{A.~Yu.~Barnyakov} \binp\nsu
\author{E.~V.~Bedarev} \binp\nsu\
\author{K.~I.~Beloborodov} \binp\nsu
\author{A.~V.~Berdyugin} \binp\nsu
\author{A.~G.~Bogdanchikov}\binp
\author{A.~A.~Botov}\binp
\author{T.~V.~Dimova}\binp\nsu
\author{V.~P.~Druzhinin}\binp\nsu
\author{R.~A.~Efremov}\binp
\author{V.~N.~Zhabin}\binp\nsu
\author{V.~V.~Zhulanov}\binp
\author{P.~V.~Zhulanova}\binp
\author{L.~V.~Kardapoltsev}\binp\nsu
\author{A.~S.~Kasaev}\binp
\author{A.~A.~Kattsin}\binp
\author{D.~P.~Kovrizhin} \binp
\author{I.~A.~Koop}\binp\nsu
\author{A.~A.~Korol}\binp\nsu
\author{A.~S.~Kupich} \binp\nsu
\author{A.~P.~Kryukov} \binp
\author{N.~A.~Melnikova} \binp
\author{N.~Yu.~Muchnoi} \binp\nsu
\author{A.~E.~Obrazovsky} \binp
\author{A.~A.~Oorzhak} \binp\nsu
\author{I.~V.~Ovtin} \binp
\author{E.~V.~Pakhtusova} \binp
\author{I.~A.~Polomoshnov} \binp
\author{K.~V.~Pugachev} \binp\nsu
\author{S.~A.~Rastigeev} \binp
\author{Yu.~A.~Rogovsky} \binp\nsu
\author{A.~I.~Senchenko} \binp
\author{S.~I.~Serednyakov} \email{S.I.Serednyakov@inp.nsk.su} \binp\nsu
\author{Z.~K.~Silagadze} \binp\nsu
\author{K.~D.~Sungurov} \binp\nsu
\author{I.~K.~Surin} \binp
\author{Yu.~V.~Usov} \binp
\author{A.~G.~Kharlamov}\binp\nsu
\author{D.~E.~Chistyakov} \binp\nsu
\author{D.~A.~Shtol} \binp

\date{ }

\begin{abstract}
In the experiment to study the  $e^+e^-\to n\bar{n}$ process at the
VEPP-2000 $e^+e^-$ collider, the ratio $|G_E|/|G_M|$ of the neutron
timelike electric and magnetic form factors has been measured. 
The experiment was carried out with the SND detector in the
center-of-mass energy range 1890-2000 MeV in eight  energy points with an integrated
luminosity of 83 pb$^{-1}$. The $|G_E|/|G_M|$ ratio is determined by the
analyzing the distribution of the polar angle of the produced antineutron.
The measured $|G_E|/|G_M|$ value in the energy range under study is
between 1.0 and 1.5 with an average value of 1.21$\pm$0.13.
\end{abstract}


\maketitle

\section*{INTRODUCTION\label{sec:intro}}
 The most important parameters of nucleons are their electromagnetic
form factors, determined by their interaction with photons. In
$e^+e^-$ annihilation the  process of  neutron-antineutron pair
production 
\begin{eqnarray}
e^+e^-\to n \bar{n}
\label{eqB0}
\end{eqnarray}
occurs through a virtual photon.  Its cross section depends
on two form factors, electric ($G_E$) and magnetic ($G_M$):
\begin{eqnarray}
\frac{d\sigma}{d\Omega}&=&\frac{\alpha^{2}\beta}{4s}
\bigg[ |G_M(s)|^{2}(1+\cos^2\theta)\nonumber\\
&+&\frac{1}{\gamma^2}|G_E(s)|^{2}\sin^2\theta
\bigg],
\label{eqB1}
\end{eqnarray}
where $\alpha$ is the fine structure constant, 
$s=4E_b^2=E^2$, where $E_b$ is the beam energy and $E=2E_b$ is the
center-of-mass (c.m.)  energy,  $\beta = \sqrt{1-4m_n^2/s}$, $\gamma
= E_b/m_n$, $m_n$ is the neutron mass, and $\theta$ is the antineutron 
production polar angle.

The total cross section has the following form:
\begin{equation}
\sigma(s) =
\frac{4\pi\alpha^{2}\beta}{3s}(1+\frac{1}{2\gamma^2})|F(s)|^2,
\label{eqB2}
\end{equation}
where   the effective form factor  $F(s)$ is introduced:  
\begin{equation}
|F(s)|^2=\frac{2\gamma^2|G_M(s)|^2+|G_E(s)|^2}{2\gamma^2 +1 }.
\label{eqB3}
\end{equation}
The effective form factor $|F(s)|^2$ shows the difference between the total cross
section (\ref{eqB2}) and the cross section for pointlike nucleons, for
which $|F(s)|^2=1$.  There is an additional condition 
$|G_{E}| = |G_{M}|$ at the threshold.

  The $e^+e^-\to n\bar{n}$ process  
was first observed in the FENICE ~\cite{FENICE} experiment. 
Recent measurements of this process
were carried out at the VEPP-2000 collider with the SND detector at
c.m. energies from the very threshold up to 2 GeV
~\cite{Art1719,Art2023,Art2023a,Art2024}.
At energies above 2 GeV, cross section data were 
obtained by the BESIII ~\cite{BES}. 
In this work we present the SND results on the $|G_E|/|G_M|$ ratio measurements 
in the energy range 1890-2000 MeV.

\begin{figure*}
\centering
\includegraphics [width = 0.7\textwidth]{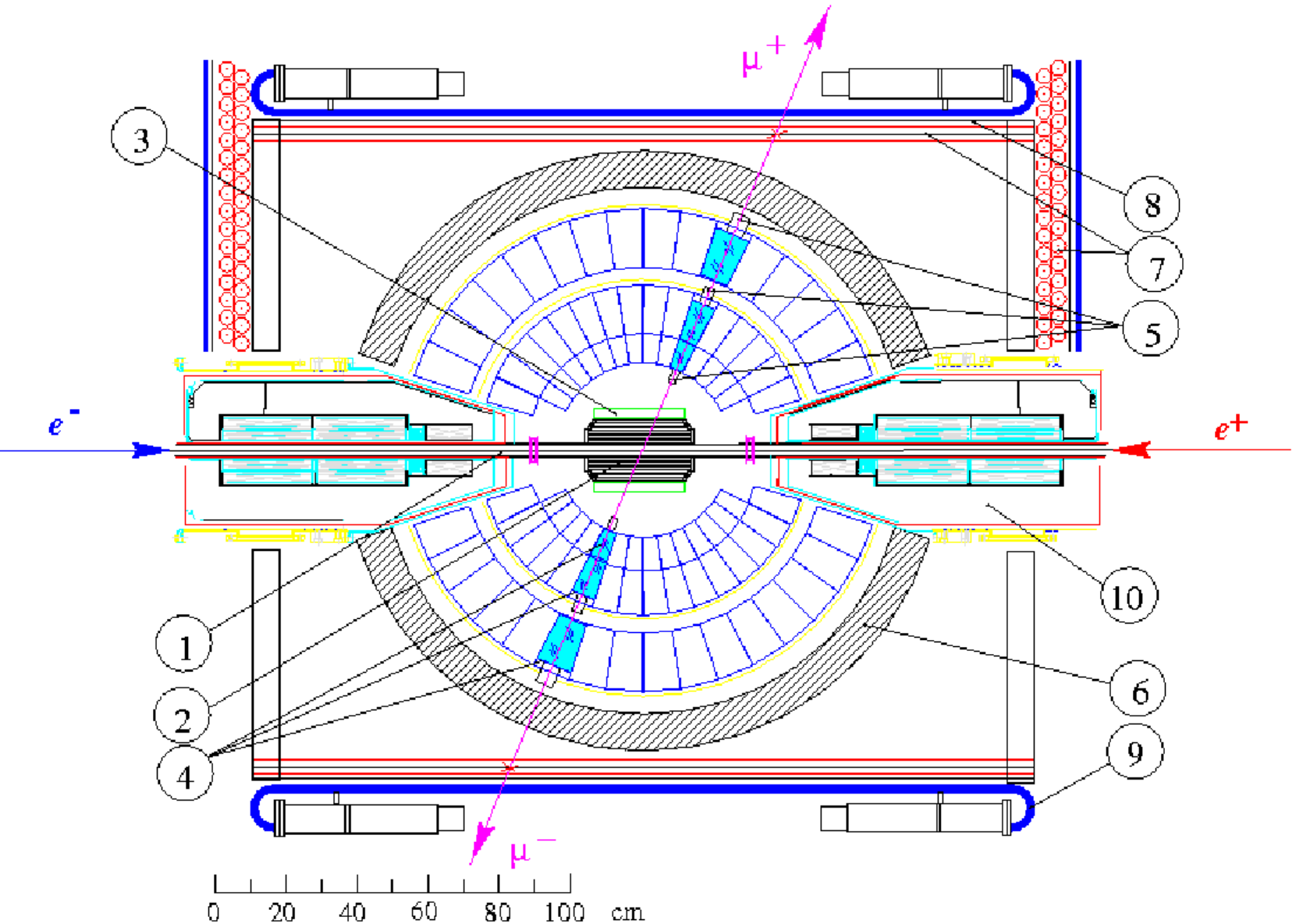}
\caption{SND detector, section along the beams: (1) beam pipe,
(2) tracking system, (3) aerogel Cherenkov counters, (4) NaI (Tl)
crystals, (5) vacuum phototriodes, (6) iron absorber, (7) proportional
tubes,
(8) iron absorber, (9) scintillation counters, (10) VEPP-2000 focusing
solenoids.}
\label{fig:sndt}
\end{figure*}

\section{COLLIDER, DETECTOR, EXPERIMENT\label{sec:Exper}}
The VEPP-2000 $e^+e^-$ collider~\cite{VEPP2k} has been operating since
year 2011. It covers the energy range from the hadron production
threshold E=280 MeV to 2 GeV. The collider luminosity depends on
energy and is about $5\times 10^{31}$~cm$^{-2}$s$^{-1}$ above
the nucleon threshold. There are two beam collision areas at  VEPP-2000,
one of which is occupied by the SND detector. 

    The SND detector~\cite{SNDet}  (Fig.\ref{fig:sndt}) and its 
operation in  $n\bar{n}$   experiments    are  described in previous 
publications~\cite{Art1719,Art2023,Art2023a}. Its main part is a
well-segmented ($\Delta\phi=\Delta\theta=9^\circ$) NaI(Tl) calorimeter (EMC).
The calorimeter thickness of 35 cm NaI(Tl)  is
sufficient to absorb neutrons and antineutrons from the process
(\ref{eqB0}). The calorimeter is equipped with the time
measurement system~\cite{Nimj} to measure event time relative to the beam
collision time. The system time resolution  measured for the
$e^+e^-\to \gamma\gamma$ events is 0.8 ns. This system plays an
important role in identifying the time-delayed $n\bar{n}$
events. The internal tracking system  is
used in anticoincidence in this analysis to suppress the
beam background. The external system (proportional tubes and scintillation
counters), located outside the EMC and the iron absorber, is used to suppress
the cosmic-ray background.

The data were collected at eight energy points in the 1890-2007 MeV range.
They are listed in Table~\ref{tab:alldat}.
The total integrated luminosity is 83 pb$^{-1}$.  The 
$e^+e^-\to n\bar{n}$ cross section in this experiment was measured 
in Ref.~\cite{Art2023a}. In this work, we measure the neutron 
$|G_E|/|G_M|$ ratio. 

\section{Event selection \label{sec:EvSelect}}
Antineutrons from the process \ref{eqB0} annihilate in EMC, giving a
large signal of up 2 GeV. The signal from the accompanying neutron is
comparatively weak and is not used. The antineutron polar angle
$\theta$, used in our analysis, is defined by the
direction of the total event momentum $P_{EMC}=\Sigma_i E_i n_i$,    
where the summation is carried out over the triggered EMC crystals, 
$E_i$ is the crystal energy deposition, $n_i$ is its position unit vector. 
Projections of the vector $P_{EMC}$ onto the direction along and across the beams 
define the polar and azimuthal antineutron angles. 

 The selection criteria for $n\bar{n}$ events were developed in our
 previous works~\cite{Art1719,Art2023,Art2023a,Art2024}. The main
 conditions are the following:
\begin{enumerate}
\item No charged tracks in the tracking system;
\item Veto signal from the external system;
\item No cosmic track or shower in EMC;
\item Large event momentum imbalance in EMC $P_{EMC}>0.4E_b$;
\item Large transverse EMC energy profile~\cite{Transv}; 
\item High total energy deposition in EMC $E_{EMC}>E_b$.
\end{enumerate}     

\begin{figure}
\includegraphics[width=0.45\textwidth]{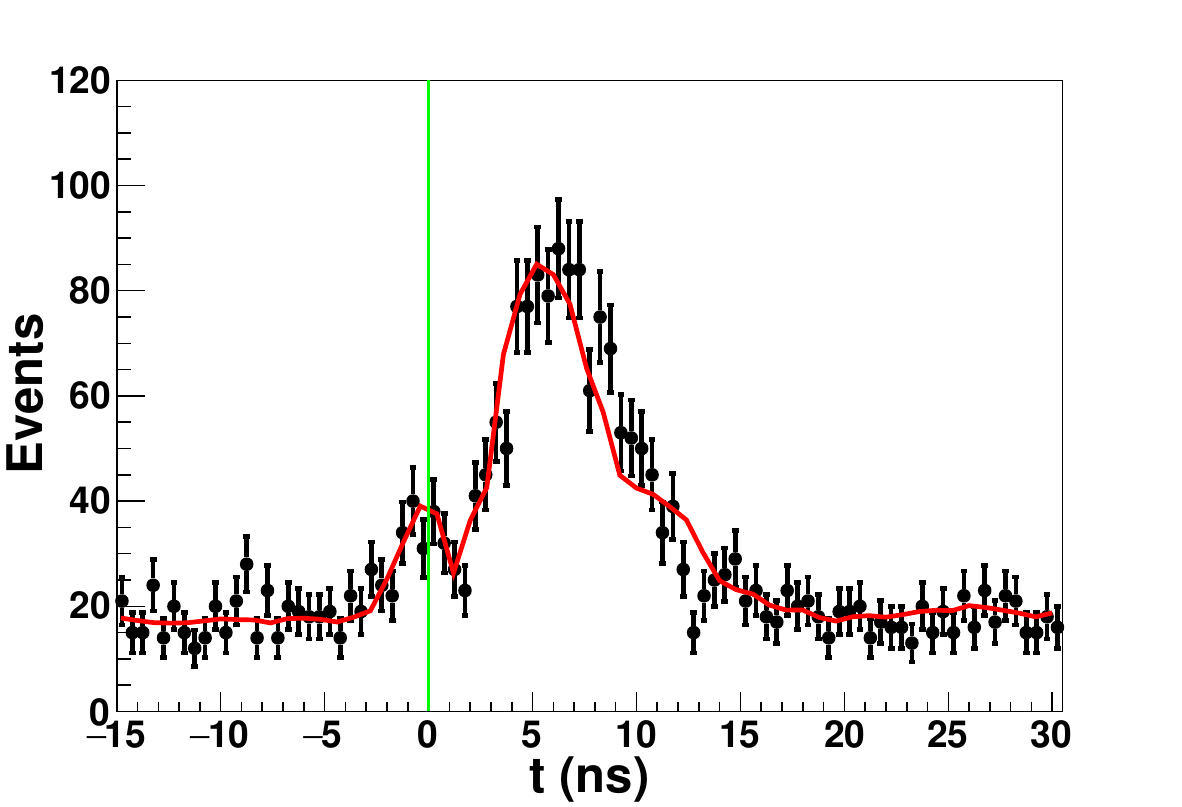} \hfill
\parbox[h]{0.45\textwidth}{\caption {The time distribution for selected
data events at $E_b=950$ MeV (points with error bars). The wide peak
to the
right is formed by delayed $n\bar{n}$ events. The vertical line at
$t=0$ shows
the time of beam collision.  The red curve is the fit of the time
spectrum described in the text.}
\label{fig:tim950}} \hfill
\end{figure}

After applying the selection criteria, approximately 400
events/pb$^{-1}$  remain for further analysis. 
The remaining events are divided into three types, depending on their
time distribution relative to the moment of the beam collision (t=0):
the cosmic-ray events, uniformly distributed in time;  the beam-induced
and physical background events concentrated around zero time, and
$n\bar{n}$ events shifted relative to zero time  by $\sim$10 ns.
An example of the time spectrum for selected events is shown in
Fig.~\ref{fig:tim950},
where these three types of events are visible. 
It is seen that the measured spectrum for $n\bar{n}$ events is quite
wide in time due to the spread of the antineutron annihilation point
inside the calorimeter.
To obtain the number of $n\bar{n}$  and background events,
the measured time spectra at each energy point are fitted by
the sum of three mentioned above contributions in the following form :

\begin{equation}
F(t)=N_{n\bar{n}}H_{n\bar{n}}(t)+N_{\rm cosm}H_{\rm
cosm}(t)+N_{\rm bkg}H_{\rm bkg}(t),
\label{timfit}
\end{equation}
where $H_{\rm cosm}$ and $H_{\rm bkg}$ are the cosmic-ray and beam background
time spectra, measured at energy below the $n\bar{n}$ threshold,
$H_{n\bar{n}}$ is the Monte Carlo (MC) simulated $n\bar{n}$ time 
spectrum~\cite{GEANT4}, $N_{n\bar{n}}$, $N_{\rm cosm}$, and $N_{\rm bkg}$
are the event numbers for the three contributions. The fitted numbers of
$n\bar{n}$ events and other data on the experiment are listed in
Table~\ref{tab:alldat}.

\begin{figure}
\includegraphics[width=0.46\textwidth]{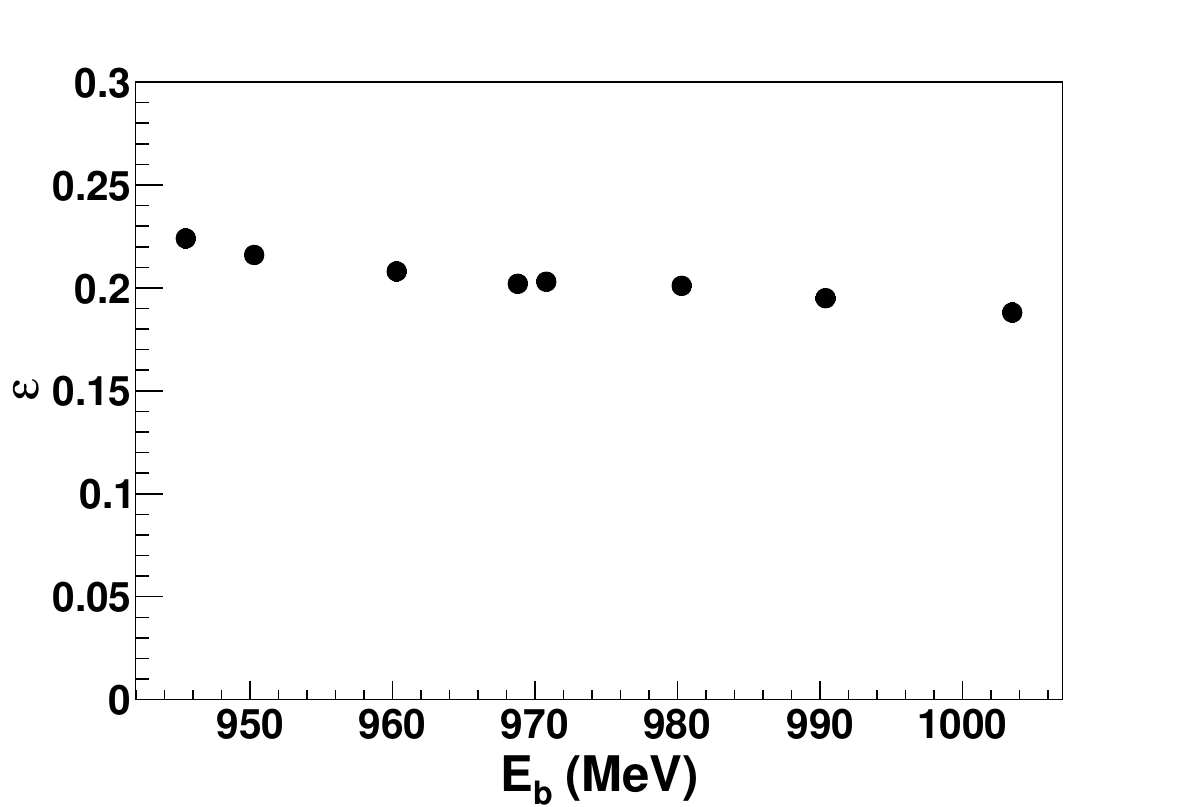} \hfill
\parbox[h]{0.46\textwidth}{\caption {
The MC detection efficiency $\varepsilon$ for $e^+e^-\to n\bar{n}$  
process versus the beam energy.}
\label{fig:efenr}} \hfill
\end{figure}

\begin{figure}
\includegraphics[width=0.46\textwidth]{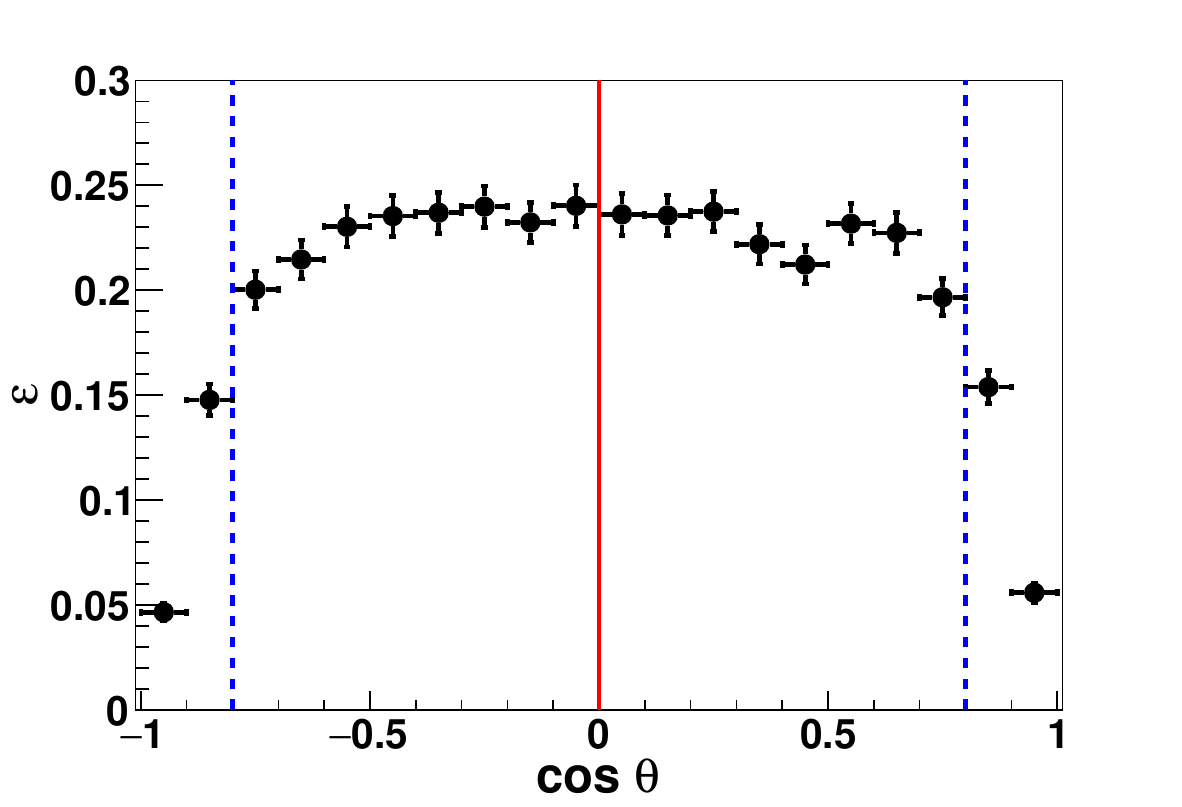}
\hfill \parbox[h]{0.46\textwidth}{\caption { The MC detection efficiency
$\varepsilon$  versus $\cos\theta$.
The vertical lines at $\cos\theta=\pm 0.8$ show the area used in
analysis. } \label{fig:effcos}}
\end{figure}

The MC detection efficiency, obtained using the selection criteria
described above, is about 20\%. It is shown in
Fig.~\ref{fig:efenr} as a function of the beam energy and in
Fig.~\ref{fig:effcos} as a function of the cosine of the
antineutron polar angle ($\cos\theta$) at the beam energy 980 MeV.

\begin{table*}
\centering
\caption{The beam energy ($E_b$), neutron
momentum ($p$),  integrated luminosity ($L$), number of selected
$n\bar{n}$ events ($N_{n\bar{n}}$), and measured $|G_E|/|G_M|$ ratio. The
quoted errors for $|G_E|/|G_M|$ and $N_{n\bar{n}}$ are  statistical and
systematic.
\label{tab:alldat}}
\begin{tabular}{cccccc c}  N & $E_b$(MeV) &
p(MeV/c) &$L$(1/pb)& $N_{n\bar{n}}$ &   $|G_E|/|G_M|$ \\ \hline
 1 & 945.5   &  51 & 9.38  & $882\pm 41$   &$1.321\pm0.401\pm0.099$ \\
 2 & 950.3   &  67 & 9.44  & $978\pm 40$   &$1.473\pm0.388\pm0.184$ \\
 3 & 960.3   &  86 & 9.17  & $962\pm 38$   &$1.371\pm0.332\pm0.225$ \\
 4 & 968.8   & 101 & 8.78  & $909\pm 38$   &$1.122\pm0.293\pm0.139$ \\
 5 & 970.8   & 122 & 5.99  & $625\pm 37$   &$0.523\pm0.440\pm0.203$ \\
 6 & 980.3   & 133 & 8.17  & $764\pm 37$   &$1.064\pm0.337\pm0.151$ \\
 7 & 990.4   & 140 & 9.40  & $811\pm 39$   &$1.352\pm0.398\pm0.306$ \\
 8 & 1003.5  & 147 & 22.5  & $1562\pm61$   &$1.530\pm0.340\pm0.115$ \\
\hline
\end{tabular}
\end{table*}
	 
\section{\label{sec:Costh}Distribution of $\cos\theta$ }%
\begin{figure}
\includegraphics[width=0.46\textwidth]{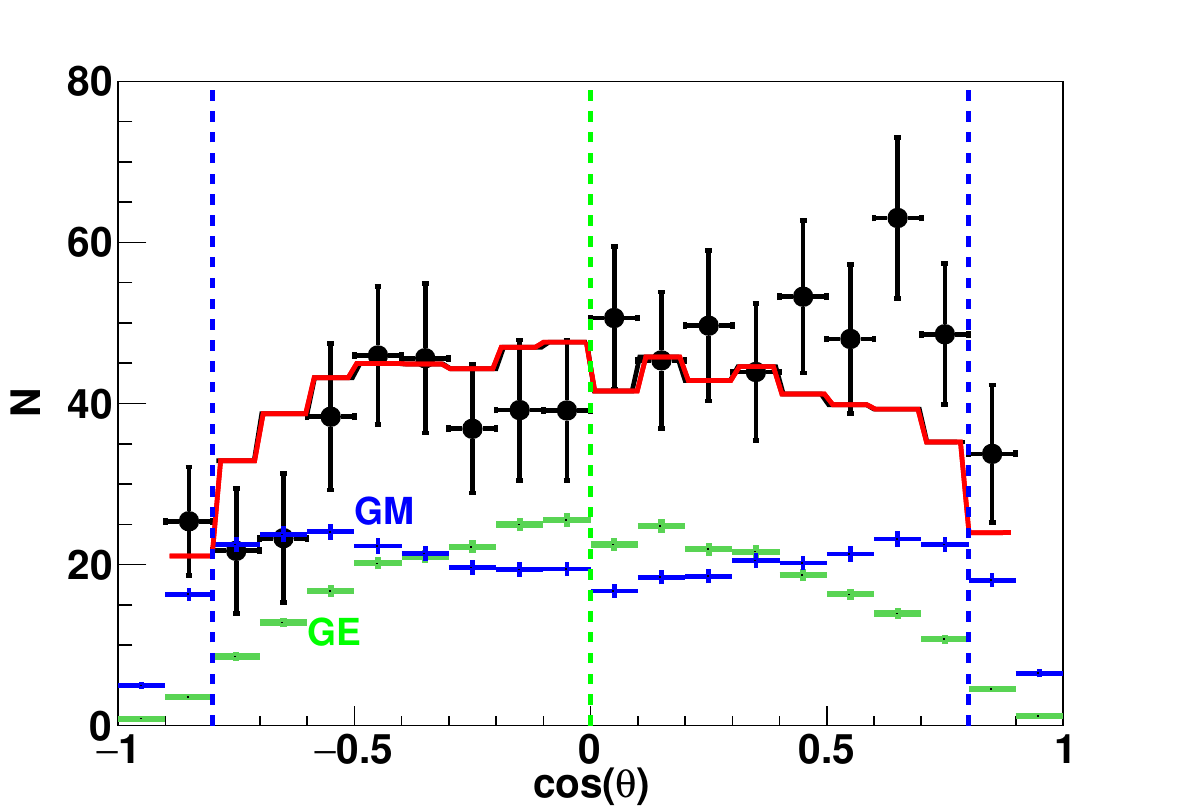}
\parbox[h]{0.48\textwidth}{
\caption {The $\cos\theta$
distribution of selected $n\bar{n}$ events at the beam energy 990
MeV (points with error bars).
The solid (red) curve is the result of the fit described in the text.
The histograms labeled $G_E$ and $G_M$
show the contributions of the electric and magnetic form factors,
respectively.  The vertical lines at $\cos\theta=\pm 0.8$
delimit the area for analysis.}  
\label{fig:cos970}}
\end{figure}

To obtain experimental $\cos\theta$ distribution, we divide the 
$-0.9<\cos\theta<0.9$ range into 18 intervals. In each 
$\cos\theta$ interval the fit to the time spectrum is performed as is
described in Sec.~\ref{sec:EvSelect} and the number of $n\bar{n}$
events is determined. The obtained $\cos\theta$ distribution at
$E_b=990$ MeV is shown in  Fig.\ref{fig:cos970}.
The bins with $|\cos\theta|=0.8-0.9$ correspond to the edge
of the sensitive area of the calorimeter. Thus, 
the main $\cos\theta$ interval, 
used in the subsequent analysis, lies within -0.8  +0.8.

\begin{figure}
\includegraphics[width=0.46\textwidth]{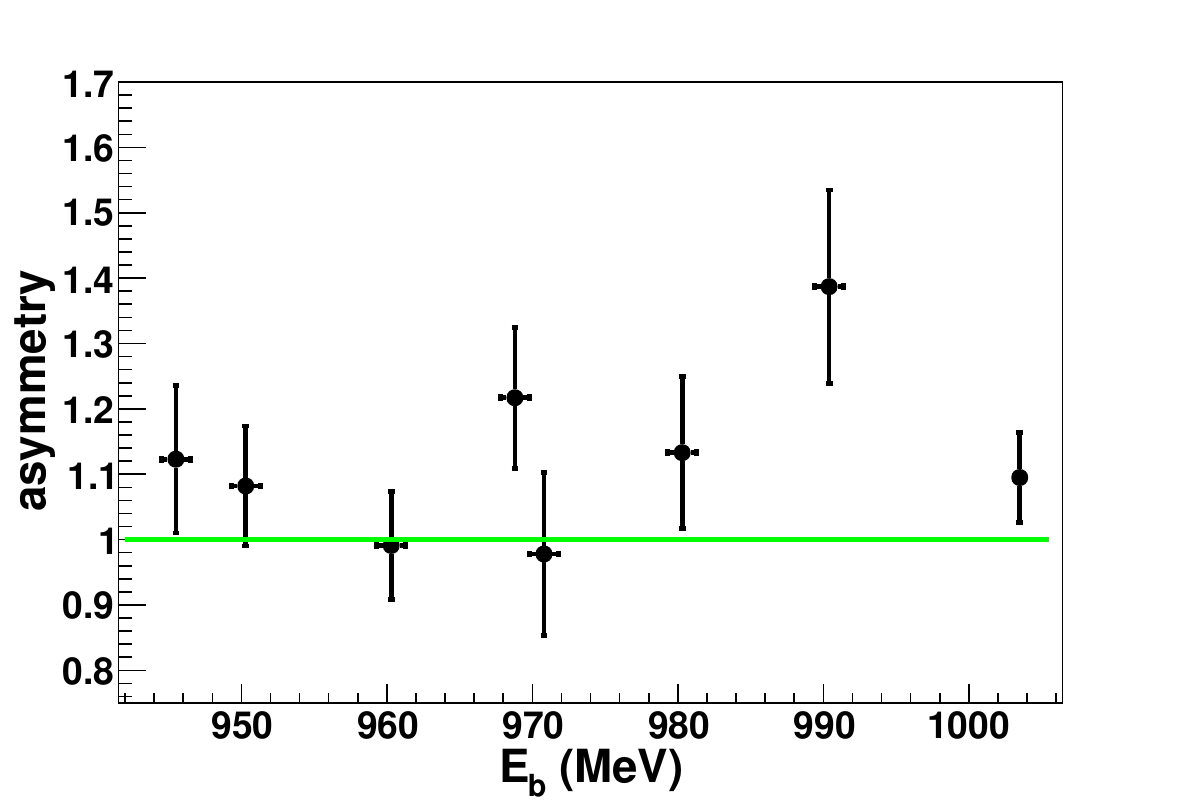}
\parbox[h]{0.46\textwidth}{\caption { The right/left asymmetry in
 the $\cos\theta$ distribution for data events as a function of the
 beam energy. The horizontal line corresponds to the case of no
 asymmetry. }
\label{fig:asym}} \hfill
\end{figure}

   When studying the data $\cos\theta$ distributions, a right/left asymmetry was
unexpectedly revealed in these spectra. It is most clearly seen in the
distribution at $E_b=990$ MeV shown in Fig.~\ref{fig:cos970}.
Since the measured angle $\theta$ corresponds
to the antineutron angle, the right/left asymmetry is also a charge
asymmetry. The asymmetry $\alpha_{as}$  is defined as the 
ratio 
$N_{n\bar{n}}(0<\cos\theta<0.8)/N_{n\bar{n}}(-0.8<\cos\theta<0)$.
The asymmetry $\alpha_{as}$ is shown in Fig.~\ref{fig:asym} as a
function of the beam energy $E_b$. Its average value is $\alpha_{as}
=1.11\pm0.04$. 
It is unlikely that this is an asymmetry of the detector itself. since
there are no any angular distortions  in other precisely 
measured processes such as $e^+e^-\to \gamma\gamma, e^+e^-  ,\pi^+ \pi^- $.
But on the other hand, we do not completely rule out 
the detector asymmetry for the 
process  $e^+e^-\to n\bar{n}$. The thing is, that for $n\bar{n}$
events there is a significant asymmetry in the energy release 
in the calorimeter: large energy up to 2 GeV from antineutron 
annihilation and small $\sim$ 10 MeV in opposite direction from neutron. 
This momentum imbalance  strongly distinguishes the
process $e^+e^-\to n\bar{n}$ from  other studied processes with
angularly symmetrical energy deposition. Perhaps the detector asymmetry 
occurs only for the $n\bar{n}$ type of events.
In the following, this asymmetry is taken into account as an additional
source of systematic error in the value of $|G_E|/|G_M|$.

  There are also physical reasons for charge asymmetry. 
It is known that charge asymmetry in the production of fermion pairs 
(e.g. muons~\cite{asmu}) can reach more than a few  percent when 
taking into account higher order QED corrections. 
In our case of $n\bar{n}$ pairs this could be, for example, a process 
$e^+e^- \to \gamma^{\star}\gamma^{\star} \to n \bar{n}$.  
In PDG Tables~\cite{pdg} there is a tensor 
state $f_2(1950)$,  decaying into a nucleon-antinucleon pair, which can 
contribute to the asymmetry due to interference of  one-photon and
two-photon intermediate channels.
But there are no calculations of charge asymmetry 
for nucleon pairs  in the
threshold region. We will facilitate  the calculations of
asymmetry in this region.

\section{Measurement of the neutron $|G_E|/|G_M|$ ratio \label{sec:gegm} }
   To obtain the $|G_E|/|G_M|$ value, the data $\cos\theta$ distribution 
shown in Fig.~\ref{fig:cos970} is fitted with the following function:   
\begin{equation}
F(\cos\theta)=N(H_M(\cos\theta)+\frac{1}{\gamma^2}|\frac{G_E}{G_M}|^2
H_E(\cos\theta)),
\label{costfit}
\end{equation}
where $H_M$ and $H_E$ are the $\cos\theta$ distributions for selected
simulated $n\bar{n}$ events generated with the angular distributions 
$1+\cos^2\theta$ and $\sin^2\theta$ (see Eq.~(\ref{eqB1})),
respectively. The shape $H_M$ and $H_E$ distributions differ from the generated
initial distributions due to nonuniform detection eﬃciency (see
Fig.~\ref{fig:effcos}), the ﬁnite $\theta$ resolution ($\sigma_\theta\approx
8^\circ$), and the radiative corrections.
The number of events $N$ and $|G_E|/|G_M|$ are free fit
parameters in Eq.~(\ref{costfit}). The result of the fit is shown in
Fig.~\ref{fig:cos970}. The fitting procedure is performed in each of 
eight  experimental energy points in Table~\ref{tab:alldat}. 

To estimate systematic uncertainties, we vary parameters that can 
influence the shape of the $\cos\theta$ distribution.
Removing the condition of no charged tracks in an event results in 
a shift in $|G_E|/|G_M|$ by an average of 0.05.
Relaxing the condition on the transverse EMC energy profile
changes $|G_E|/|G_M|$ by approximately 0.1. 
Variations of the time position of the beam background peak within 0.5 ns
lead to a significant effect at $E_b>700$ MeV. The change in $|G_E|/|G_M|$ 
is about 0.08.
To check the influence of the asymmetry in the $\cos\theta$ distribution,
we fit the $|\cos\theta|$ distribution. The obtained shift in
$|G_E|/|G_M|$ is about 0.06 at $E_b<700$ MeV and about 0.11 at $E_b>700$ MeV. The
systematic uncertainties from all sources are combined quadratically.
The obtained $|G_E|/|G_M|$ values with statistical and systematic
uncertainties are listed in Table~\ref{tab:alldat}.

\begin{figure}
\includegraphics[width=0.46\textwidth]{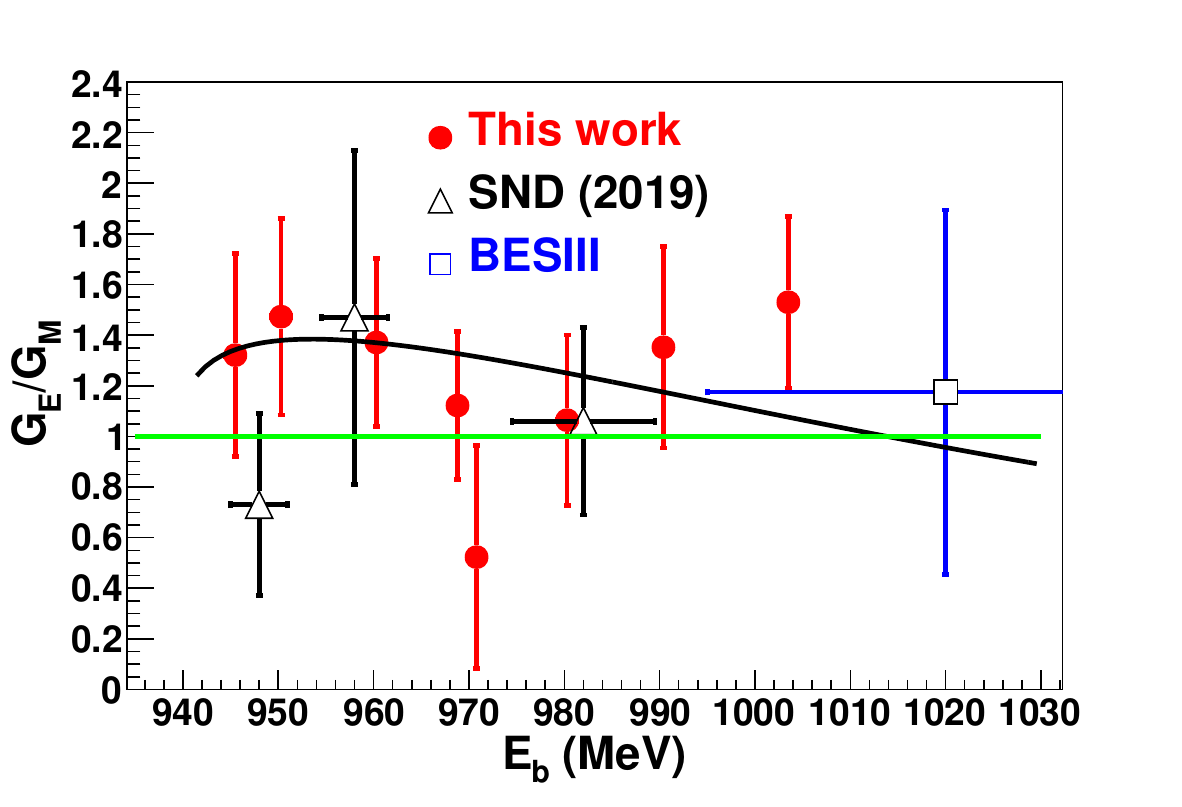} \hfill
\parbox[h]{0.46\textwidth}{\caption {  The dependence of the measured
$|G_E|/|G_M|$ ratio (solid circles) on the beam energy.  Data  of
earlier measurements SND~\cite{Art1719} (empty triangle) and BESIII~\cite{BES}
(solid stars) are shown as well. The horizontal line corresponds to the 
threshold value  $|G_E|/|G_M|=1$. Solid curve is the prediction 
from Ref.~\cite{theor}. } 
\label{fig:gemeb}} 
\end{figure}

\begin{figure}
\includegraphics[width=0.46\textwidth]{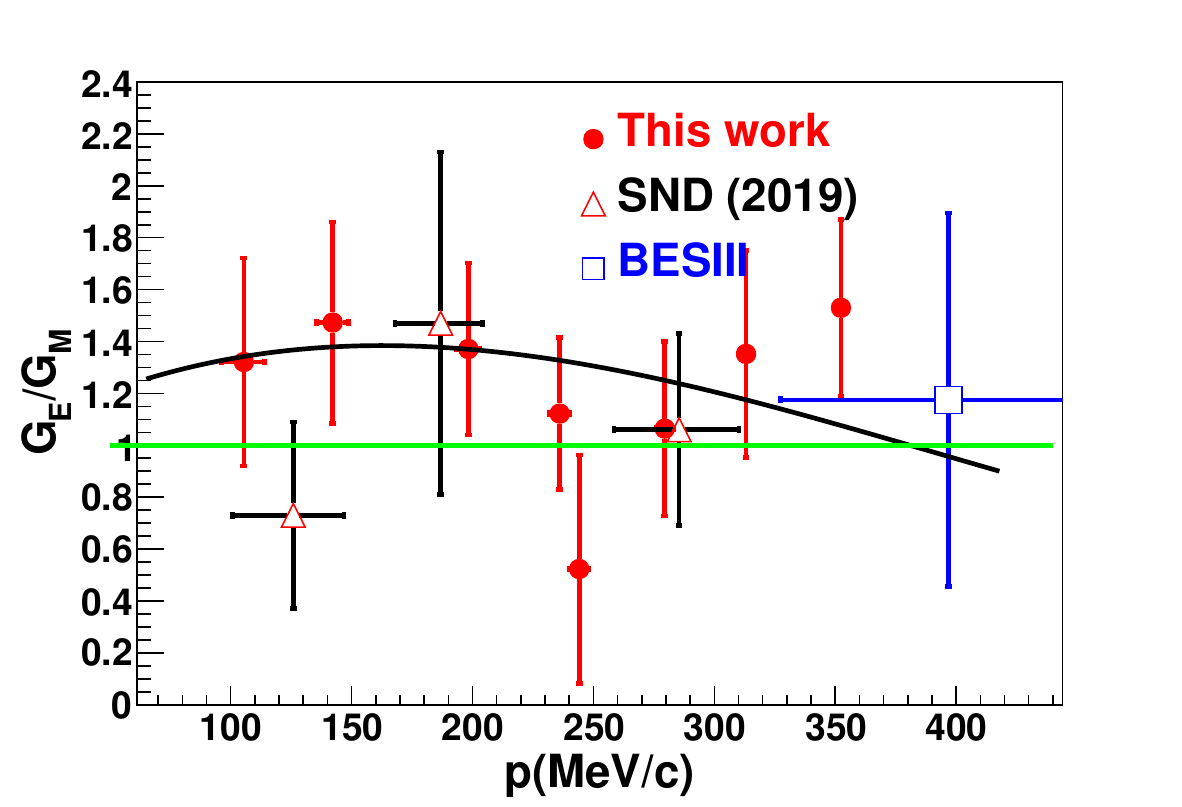}
\parbox[h]{0.46\textwidth}{\caption {The dependence of the measured
$|G_E|/|G_M|$ ratio (solid circles) on the antineutron momentum.
Solid curve is the prediction from Ref.~\cite{theor}.}
\label{fig:gempb}} 
\end{figure}

    The measured values of the $|G_E|/|G_M|$ ratio versus the beam
energy are shown in Fig.\ref{fig:gemeb}. The $|G_E|/|G_M|$
dependence on the antineutron momentum is presented in
Fig.\ref{fig:gempb}. Existing data of
experiments~\cite{Art1719,BES} are also shown in both plots for
comparison. We conclude that the measured $|G_E|/|G_M|$ values lie
mainly within the range 1.0--1.5. The obtained results are in
moderate agreement with the calculations of Ref.~\cite{theor}. 

\section{Summary \label{sec:summ} }     
   In the experiment to study the $e^+e^-\to n\bar{n}$ process, 
the ratio $|G_E|/|G_M|$ of the neutron electric and magnetic form
factors in timelike region has been  measured. The experiment was
carried out at the VEPP-2000 $e^+e^-$ collider with the SND detector
at eight  energy points in the beam energy range from 945 to 1000 MeV. The
measured value of $|G_E|/|G_M|$ lies within 1.0 - 1.5. The statistical
accuracy of measurements is about 0.30, while the systematic
uncertainties are in the range 0.1 - 0.3. The obtained data do not
contradict earlier measurements of SND~\cite{Art1719} and
BESIII~\cite{BES}, as well as the theoretical calculation~\cite{theor}.

        \bf{ACKNOWLEDGEMENTS}

The authors express their gratitude to A.I. Milstein for useful
discussions of theoretical aspects of nucleon form factors.  This work
was partially supported by the Grant No. 23-22-00011 of the Russian
Science Foundation.

      \bf{DATA AVAILABILITY}
	
The data that support the findings of this article are not publicly	
available and/or the cost of preparing, depositing, and hosting the
data would be prohibitive within the terms of this research project.
The data are available from the authors upon reasonable request.

\end{document}